\documentclass[11pt]{article}
\usepackage{amsmath}
\usepackage{graphicx}
\usepackage{bm}
\usepackage{subfigure}
\usepackage[margin=15pt,font=small,labelfont=bf]{caption}
\usepackage{lineno}
\linenumbers
\begin{document}

\title{Accumulation and extraction of ultracold neutrons from a superfluid
helium converter coated with fluorinated grease}
\author{O. Zimmer$^{1,2\ast }$, P. Schmidt-Wellenburg$^{1,2}$, M. Assmann$%
^{1}$, M. Fertl$^{1}$, \and J. Klenke$^{3}$, S. Mironov$^{1,4}$, H.-F. Wirth$%
^{1}$, B. van den Brandt$^{5}\bigskip $ \\
$^{1}$Physik-Department E18, Technische Universit\"{a}t M\"{u}nchen, \\
D-85748 Garching, Germany\\
$^{2}$Institut Laue-Langevin, B.P. 156, 38042 Grenoble, France\\
$^{3}$Forschungsreaktor M\"{u}nchen FRM II, Lichtenbergstrasse 1, 85747
Garching, Germany\\
$^{4}$Laboratory of Nuclear Problems, JINR, Dubna, Moscow region 141980,
Russia\\
$^{5}$Paul Scherrer Institut, CH-5232 Villigen PSI, Switzerland}
\maketitle

\begin{abstract}
We report experiments on the production of ultracold neutrons (UCN) in a
converter of superfluid helium coated with fluorinated grease. We employed
our technique of window-free extraction of accumulated UCN from the helium,
in which they were produced by downscattering neutrons of a cold beam from
the Munich research reactor. The time constant for UCN passage through the
same extraction aperture as in a previous experiment was a factor two
shorter, despite a lower mean velocity of the accumulated UCN in the present
experiments. A time-of-flight measurement of the cold neutron spectrum
incident on the converter allowed us to estimate the multi-phonon
contribution to the UCN production. The UCN production rate inferred from
two methods agrees with the theoretical expectation.\bigskip

\medskip PACS numbers: 78.70.Nx, 28.20.Fc, 29.25.Dz, 61.12.Ha

Keywords: ultracold neutrons, UCN, UCN sources\bigskip

$^{\ast }$email: zimmer@ill.fr
\end{abstract}

\section{Introduction}

Ultracold neutrons (UCN) have energies in the neV range and velocities up to
a few meters per second. When impinging on suitable materials they undergo
total reflection under any angle of incidence and can therefore be trapped
in bottles and manipulated for a long time (see the books \cite%
{Golub/1991,Ignatovich/1990} for an introduction to the physics of UCN).
Owing to this feature UCN have become very useful in various fundamental
investigations of neutron properties, with strong implications for particle
physics and cosmology. The longstanding search for the neutron electric
dipole moment investigates CP-violation beyond the standard model of
particle physics \cite{Baker/2006,Pospelov/2005}. The accurate determination
of the neutron lifetime is required for a detailed understanding of big bang
nucleosynthesis \cite{Lopez/1999}, and to investigate strength and structure
of the semi-leptonic weak interaction within the first quark family (see,
e.g., the workshop proceedings \cite{Arif/2005,Abele/2002}). New topics with
UCN are, among others, the demonstration of quantum levels of neutrons in
the earth's gravitational field \cite{Nesvizhevsky/2003}, and even more
recently, the search for neutron - mirror neutron transitions \cite%
{Ban/2007,Serebrov/2007}. The most intense UCN source at the Institut Laue
Langevin in Grenoble \cite{Steyerl/1986} provides not more than $50$ UCN per
cm$^{3}$. In order to improve counting statistics, new UCN sources are being
developed in many laboratories around the world \cite%
{Trinks/2000,Fomin/2000,Saunders/2004,Pokot/1995,Masuda/2002,Baker/2003,LANSCE/2004}%
.

An elegant method to produce UCN employs a converter of superfluid $^{4}$He
in a beam of cold neutrons \cite{Golub/1975}. The kinematics defined by the
dispersion relations of helium and the free neutron enables downscattering
of cold neutrons with an energy around $1.0$ meV (wavelength $0.89$ nm) to
ultracold energies via emission of a single phonon. In addition,
multi-phonon processes occur which contribute to the integral UCN production
rate for a wide range of incident neutron energies \cite%
{Korobkina/2002,Schott/2003}. At low temperatures the probability for
upscattering is strongly suppressed by the Boltzmann factor. Therefore, and
since pure $^{4}$He has no cross section for neutron absorption, a large
density of UCN may build up in a converter with reflective walls. The
storage time constant $\tau $ can of course not exceed a limit close to $900$~s set by the neutron beta decay lifetime.

In a recent experiment we have demonstrated for the first time that one may
efficiently extract UCN from the converter after having them accumulated
therein \cite{Zimmer/2007}. As in past experiments \cite%
{Masuda/2002,Baker/2003,Huffman/2000,Ageron/1978} the UCN production rate
was found to agree reasonably well with the theoretical expectation. We have
developed a new cryostat designed to keep the source portable and easy to
operate. With a short cooling cycle of a few days, our system is much more
flexible than an earlier apparatus \cite{Yoshiki/1994}, which was designed
to be installed close to the target of a spallation source. Our present
apparatus is a prototype for a future UCN source to be installed at a strong
cold neutron beam of a high-flux reactor, where no extraordinary cooling
power is required. A particular feature is the vertical extraction of the
UCN through a cold mechanical valve situated above the helium bath. In
contrast to a previous attempt to extract accumulated UCN horizontally \cite%
{Kilvington/1987}, no gaps or windows are required in our method. To gain
first experience with this system, the UCN converter vessel and extraction
guide system were made of stainless steel. In the first run it enabled us to
measure, with negligible background, the UCN production rate and to study
the temperature-dependent storage properties of the converter. Here we
report new results obtained with the converter vessel coated with
fluorinated grease (Fomblin). It has good reflection properties for UCN and,
to our very surprise, positively influences the extraction time constant.
Moreover, a time-of-flight (TOF) measurement of the incident cold neutron
beam helped to improve the comparison with the theoretical UCN production
rate.

\section{Apparatus}

The apparatus was already described in some detail in ref. \cite{Zimmer/2007}%
. The converter vessel has a volume of about $2.4$ liters, made from an
electropolished stainless steel tube with length $696$ mm and inner diameter 
$66$ mm, closed by Ni windows on both ends. The lowest temperature attained
for the completely filled vessel was $T=0.72$ K in the previous run, and $%
0.82$ K in the present. The maximum kinetic energy of UCN storable in the
vessel is defined by the Fermi potential $V_{\mathrm{F,wall}}$ of the wall
material, which is $(184\pm 4)$ neV for the stainless steel used (and $252$
neV for the Ni windows), from which one has to subtract the Fermi potential
of the superfluid helium ($V_{\mathrm{F,}^{4}\mathrm{He}}=18.5$ neV). Due to
a small neutron absorbing aperture with diameter $33$ mm placed at the
entrance window to the converter, chosen in order to avoid activation of the
vessel in these first test experiments, the volume intersected by the cold
neutron beam ("UCN production volume") was $V_{\mathrm{p}}=595$ cm$^{3}$,
only. In the experiments reported here, the inner surface of the vessel was
coated with a thick layer of fluorinated grease (Fomblin) with a Fermi
potential of $(115\pm 10)$ neV. UCN were extracted as previously through a
flapper valve situated above the superfluid helium in the "T" section of the
storage tube, connecting the (uncoated) extraction line made of
electropolished stainless steel to a $^{3}$He-gas UCN detector. The present
experiments were again performed at the neutron guide "NL1" at the Munich
research reactor FRM II, using the same neutron beam collimation as in the
previous setup \cite{Zimmer/2007}.

\section{Definition of time constants and measurements}

UCN storage and extraction from the converter can be characterised by
various time constants. The storage time constant $\tau $ quantifies the
temporal decrease of UCN in the closed vessel. The rate $\tau ^{-1}$
contains a contribution $\tau _{0}^{-1}$ due to wall collisions, absorbing
impurities, and UCN escaping through small holes in the vessel. This
contribution does not depend on temperature $T$ but on the kinetic energy $E$
of the UCN. In addition there are the rates for the $T$-dependent UCN
upscattering, and for neutron beta decay\footnote{%
In the present experiments this contribution forgotten to mention in ref. 
\cite{Zimmer/2007} was still small compared to the sum of the other rates.},

\begin{equation}
	\tau ^{-1}=\tau _{0}^{-1}\left( E\right) +\tau _{\mathrm{up}}^{-1}\left(
	T\right) +\tau _{\beta }^{-1}.  
\label{tau}
\end{equation}

The emptying time constant $\tau _{\mathrm{e}}$ quantifies the temporal
decrease of UCN in the vessel with the UCN valve open. Therefore, 

\begin{equation}
	\tau _{\mathrm{e}}^{-1}\left( T,E\right) =\tau ^{-1}\left( T,E\right) +\tau_{A}^{-1}\left( E\right) ,  \label{tau-e}
\end{equation}

\noindent where $\tau _{A}$ is the time constant for UCN passage through the
extraction hole with area $A$. From these time constants we may derive the
detection probability $W$ for a UCN created in the converter vessel. It
characterises the efficiency of the whole system including extraction and is
given by

\begin{equation}
W=\varepsilon \frac{\tau _{A}^{-1}}{\tau _{\mathrm{e}}^{-1}}=\varepsilon 
\frac{\tau -\tau _{\mathrm{e}}}{\tau },  \label{W}
\end{equation}

\noindent where the factor $\varepsilon $ describes losses in the extraction line and
imperfect detector efficiency. The conversion process employed in the present experiments produces a broad
spectrum of low-energy neutrons. Only neutrons with energies below $V_{\mathrm{F,wall}}-V_{\mathrm{F,}^{4}\mathrm{He}}$ are trapped inside the
converter vessel, whereas those with higher energies quickly escape.
\begin{figure}
	\centering
	\includegraphics[]{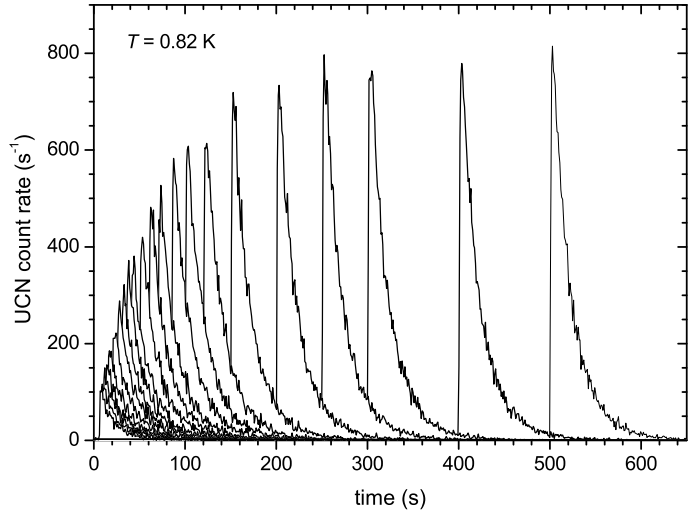}
\caption{Time histograms of UCN count rates, measured in "buildup mode" (see
text) at $0.82$ K for $23$ different accumulation times $t_{0}$.}
\end{figure}
The spectrum of the neutrons remaining in the vessel is shaped due to wall
losses. The energy dependence of these losses is due to the increase with
UCN energy of both the average loss probability per wall collision and the
frequency of wall collisions. Thus the largest losses occur for those
neutrons with energies close to the Fermi potential. However, if on a given
time scale the change of the energy spectrum stays sufficiently small, the
time constant $\tau $ is still well defined. This is found to be a
reasonably good approximation for the accumulation of UCN at high
temperatures, where the energy-independent $\tau _{\mathrm{up}}^{-1}\left(
T\right) $ dominates the rate $\tau ^{-1}$ (see eq.(\ref{tau})). At lower
temperatures we may still deduce values for the various $\tau $s (still
calling them "time constants"), and study how they depend on the times of
UCN accumulation and trapping.

\begin{figure}
	\centering
	\includegraphics[]{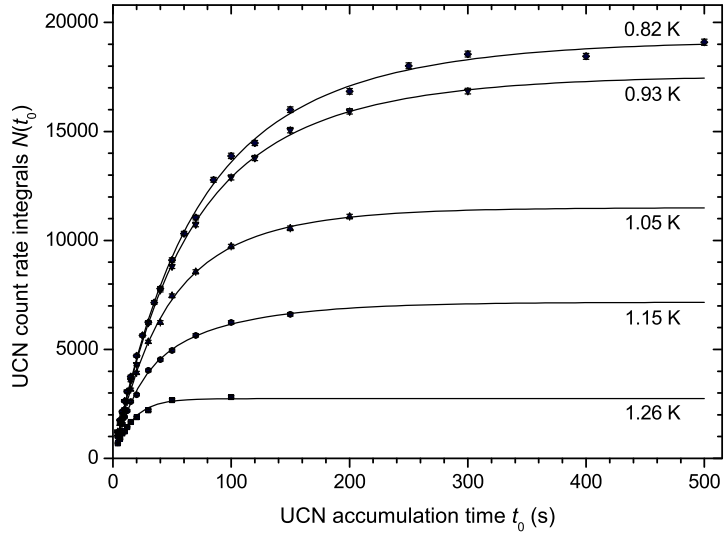}
\caption{UCN count rate integrals as a function of the accumulation time $%
t_{0}$. The fit to the data points at $T=1.26$ K is performed with the
single-exponential function defined in eq.(\protect\ref{UCN buildup}). The
results for the fitting parameters $\protect\tau $ and $N$ are employed in
section 4 to determine the UCN production rate. The fits to the data points
for lower temperatures employ double exponentials and serve as guides to the
eye.}
\end{figure}

The time constants $\tau $ and $\tau _{\mathrm{e}}$ can be obtained from
measurements in the "buildup mode". There, the closed converter is first
irradiated with cold neutrons for a time $t_{0}$, after which the beam is
shut off and simultaneously the UCN valve is opened. During the whole
process a time histogram of the UCN count rate is recorded. Figure 1 shows a
series of such histograms for different accumulation times $t_{0}$, which
was obtained for the lowest temperature of the converter in the present
runs. The decrease of the UCN count rate while emptying the vessel proceeds
with a pure single-exponential; fits of the whole decay in each of the
histograms with $A\exp (-t/\tau _{\mathrm{e}})$ always resulted in a reduced 
$\chi ^{2}$ close to unity and provided a value for $\tau _{\mathrm{e}}$ for
each of the histograms.

\begin{figure}
	\centering
	\includegraphics{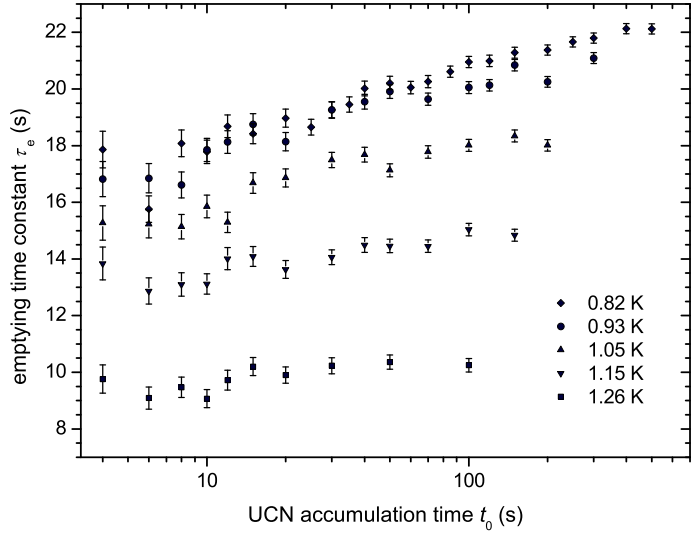}
\caption{Emptying time constants $\protect \tau _{\mathrm{e}}$ as a function
of the UCN accumulation time $t_{0}$, determined from the histograms of all
buildup-mode measurements for the various converter temperatures.}
\end{figure}

Figure 2 shows the integrals $N\left( t_{0}\right) $ of all histograms
measured. It demonstrates the saturation behaviour and the strong $T$
dependence of UCN accumulation for the range of temperatures investigated.
If the time constant $\tau $ is sufficiently well defined, one can fit to
the data the single-exponential buildup function

\begin{equation}
N\left( t_{0}\right) =N\left( 1-\exp \left( -t_{0}/\tau \right) \right) .
\label{UCN buildup}
\end{equation}

\noindent For the highest temperature, $T=1.26$ K, this fit has a reduced $\chi ^{2}$
of $3.5$. The fit becomes increasingly worse for the lower temperatures. As
visible in fig. 3, there is also a dependence of $\tau _{\mathrm{e}}$ on $%
t_{0}$, which becomes increasingly pronounced for lower temperatures. These
facts indicate the increasing influence of the first term in eq.(\ref{tau})
with decreasing temperature. The fit of a constant value to the data for $%
\tau _{\mathrm{e}}$ at $1.26$ K has $\chi ^{2}$ $=2.3$. For later use we
summarise below the results of the fits to the data at this temperature, and
the corresponding value for $\tau _{A}$, obtained with eq.(\ref{tau-e}) and
the value for $\tau $ from eq.(\ref{fitting parameters}). The uncertainties
stated for $\tau $, $N$, and $\tau _{\mathrm{e}}$ have been scaled to
provide a fit with $\chi ^{2}=1$:

\begin{eqnarray}
\tau  &=&(15.79\pm 0.82)\text{ s},  \nonumber \\
N &=&2740\pm 73,  \label{fitting parameters} \\
\tau _{\mathrm{e}} &=&(9.92\pm 0.15)\text{ s,}  \nonumber \\
\tau _{A} &=&(26.7\pm 2.6)\text{ s.}  \nonumber
\end{eqnarray}

\begin{figure}[ptb]
	\centering
	\includegraphics[]{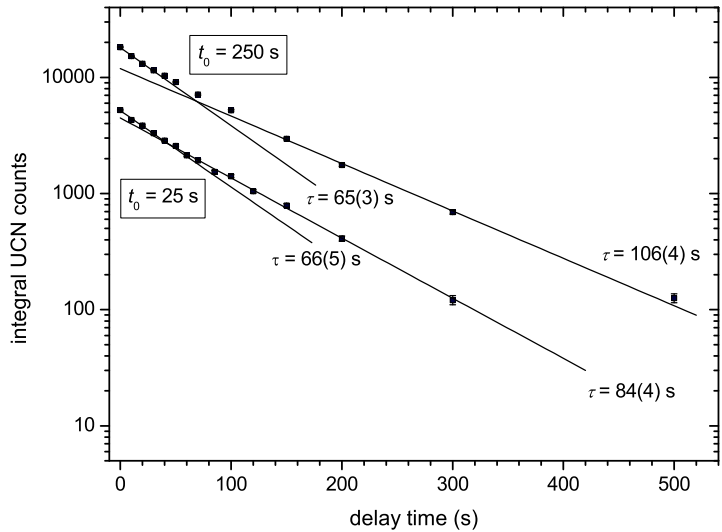}
\caption{Count rate integrals of the measurements at $0.82$ K with delayed
extraction after UCN accumulation times $t_{0}=25$ s (lower data points) and 
$250$ s (upper data points). The solid lines are fits of single exponentials
to the four first, respectively, four last data points. The corresponding
time constants $\protect\tau $ derived from these fits are also shown.}
\end{figure}

From the buildup-mode measurements as described before we cannot extract
values for $\tau $ as a function of the time the UCN stay trapped in the
vessel. In order to demonstrate this dependence we employed a variant of
such measurements with a delayed extraction. There again the converter is
first irradiated with cold neutrons while the UCN valve stays closed. After
an accumulation time $t_{0}$ the beam is shut, but the UCN valve is opened
only after a delay time $t_{\mathrm{d}}$. A series of such measurements
provides histograms for the same $t_{0}$ but various $t_{\mathrm{d}}$. A
plot of the count rate integrals of two series, for $t_{0}=25$ s and $250$
s, is shown in fig. 4. Values for $\tau $ can be obtained from
single-exponential fits to a group of several data points. The variation of
slope of the curves shown in fig. 4 indicates a more pronounced increase of $\tau $ with $t_{\mathrm{d}}$ when UCN are accumulated for a longer time.
This is to be expected, since after $t_{0}=250$ s the relative abundance of
slower UCN with respect to the (more abundant) faster ones will be
significantly higher than after $25$ s, due to the longer storage time
constant of the slower UCN. Their presence in the vessel then becomes better
visible in the delayed-extraction experiments.

Values for $\tau $ can also be deduced from the integral UCN counts for only
two different delay times, using the relation

\begin{equation}
\tau =\frac{t_{\mathrm{d}_{2}}-t_{\mathrm{d}_{1}}}{\ln N\left( t_{\mathrm{d}%
_{1}}\right) -\ln N\left( t_{\mathrm{d}_{2}}\right) }.  \label{tau storage}
\end{equation}

\noindent Figure 5 shows such an analysis for the series with $t_{0}=250$ s. The
strong variation of time constants from one to two minutes demonstrates
indeed how much at $0.82$ K the spectrum of the UCN remaining in the vessel
is shaped by the energy dependence of the wall collisions during trapping.

\begin{figure}[ptb]
	\centering
	\includegraphics[]{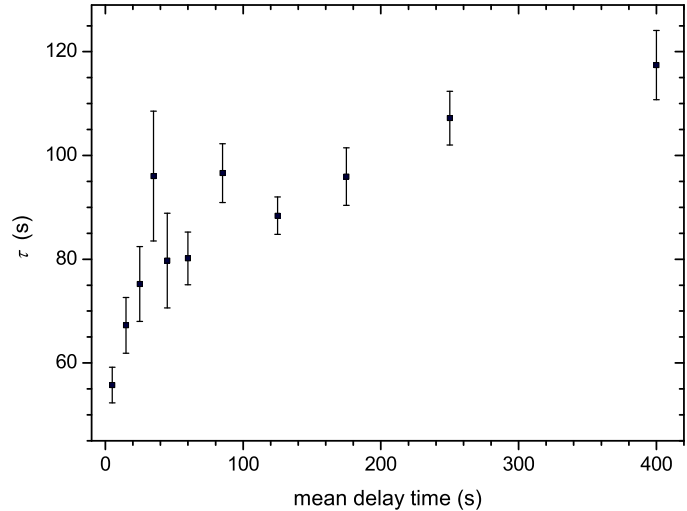}
\caption{UCN storage time constants deduced from eq.(\protect\ref{tau
storage}), using adjacent data points from the measurements with delayed
extraction shown in fig. 4. The abscissa is given by the mean of the two
adjacent delay times, $(t_{\mathrm{d}_{i+1}}+t_{\mathrm{d}_{i}})/2$.}
\end{figure}

In a third type of measurements, called "continuous mode", we measured the
steady state count rate $\dot{N}_{\mathrm{c}}$ with the converter irradiated
for a long time with cold neutrons while the UCN valve stayed open. The
count rates measured are listed in Table 1. In the next section we will use $ \dot{N}_{\mathrm{c}}$ and $W/\varepsilon $ at $1.26$ K as input to determine
the UCN production rate. For the purpose of illustration we present in Table
1 also values for $W/\varepsilon $ derived from $W(T_{i})=W(1.26\,$K$)\dot{N}_{\mathrm{c}}(T_{i})/\dot{N}_{\mathrm{c}}(1.26$\thinspace K$)$, which
assumes that the same UCN spectrum prevails for the different temperatures.
Remember that, as obvious from fig. 3, this assumption is only valid as an
approximation. The values stated for the lower temperatures might represent
the true value with a relative error in the order of $10$ \%.

\begin{table}[tbp] 
	\centering
\begin{tabular}{c|ccccc}
$T$ [K] & $1.26$ & $1.15$ & $1.05$ & $0.93$ & $0.82$ \\ \hline
$\dot{N}_{\mathrm{c}}$ [s$^{-1}$] & $176\pm 2.6$ & $227.6\pm 2.7$ & $%
257.5\pm 1.4$ & $278.3\pm 2.3$ & $286.6\pm 1.6$ \\ 
$W/\varepsilon $ [\%] & $37.1\pm 3.4$ & $48.0$ & $54.3$ & $58.7$ & $60.4$%
\end{tabular}
\caption{Measured $T$-dependent count rates obtained in continuous-mode
measurements, and values deduced for $W/\varepsilon $ (see text).}\label%
{Table1}
\end{table}

\section{UCN production rate}

The UCN production rate can be inferred from two methods, for which we
employ the data at the highest temperature, $T=1.26$ K, where the time
constants are sufficiently well defined.\\
First, as in the analysis of our last experiments \cite{Zimmer/2007}, we may
use the stationary "continuous mode" count rate $\dot{N}_{\mathrm{c}}$. The
corresponding UCN production rate density $P_{1}$ is given by

\begin{equation}
P_{1}=\frac{\dot{N}_{\mathrm{c}}}{V_{\mathrm{p}}W}.  \label{P}
\end{equation}

\noindent Using the values for $\dot{N}_{\mathrm{c}}$ and $W/\varepsilon $ stated in
Table 1 and the value $V_{\mathrm{p}}=595$ cm$^{3}$ for the UCN production
volume, we obtain

\begin{equation}
\varepsilon P_{1}=(0.797\pm 0.074)\text{ s}^{-1}\text{cm}^{-3}.  \label{P1}
\end{equation}

\noindent Lacking the knowledge of $\varepsilon $, the numerical value in eq.(\ref{P1}) provides a lower limit for $P_{1}$. \\
A second value for the production rate density, $P_{2}$, may be derived from
the saturated UCN number $N$ in the buildup-mode measurement, employing eq.(\ref{UCN buildup}). When UCN are accumulated for a long time, $t_{0}\gg \tau $, the production rate equals the UCN loss rate. Denoting with $N_{0}$ the
saturated number of UCN in the vessel, of which only the fraction $N=N_{0}W$
is detected, this leads us to

\begin{equation}
P_{2}=\frac{N}{WV_{\mathrm{p}}\tau }.
\end{equation}

\noindent Using the values for $\tau $ and $N$ from eq.(\ref{fitting parameters}) with 
$W/\varepsilon $ from Table 1 we obtain

\begin{equation}
\varepsilon P_{2}=(0.786\pm 0.085)\text{ s}^{-1}\text{cm}^{-3}.  \label{P2}
\end{equation}

\noindent Hence, within the $10$ \% accuracy of these two methods, $P_{1}=P_{2}$. We may now compare the experimental findings with the theoretical
expectation. This requires knowledge of the incident cold neutron spectrum
which we measured in a separate experiment, using a time-of-flight (TOF)
analysing device described in ref. \cite{Zeitelhack/2006}. It consists of a
mechanical chopper and a $^{3}$He detector with a vertical slit. During a
measurement the detector was moved horizontally in order to integrate over
the whole divergence of the beam. The entrance aperture of the TOF analyser
was placed at the position where previously was situated the entrance window
to the converter vessel. We thus determined the neutron wavelength spectrum
up to $\lambda =2$ nm (frame overlap of the chopped bunches occured only for 
$\lambda >5$ nm, where the intensity is negligibly small). The part for $%
\lambda \leq 1$ nm is shown in fig. 6. After the TOF measurements the
spectrum was calibrated with a gold foil activation by the integral neutron
flux.

The mechanism of UCN production contains contributions from single phonon
emission and multiphonon processes. The single-phonon contribution to the
production rate density in a helium converter with Be wall coating is $P_{\mathrm{I}}=\left( 4.55\pm 0.25\right) \times 10^{-9}\left. \text{d}\phi /\text{d}\lambda \right\vert _{\lambda ^{\ast }}$ s$^{-1}$cm$^{-3}$, where
the differential flux at $\lambda ^{\ast }=0.89$ nm is given in cm$^{-2}$s$^{-1}$nm$^{-1}$ \cite{Baker/2003}. For the present situation this value
needs to be corrected for the Fermi potential of the Fomblin grease, i.e.
divided by $\left( 252-18.5\right) ^{3/2}/\left( 115-18.5\right) ^{3/2}=3.76$, with an uncertainty of $0.59$ due to the poor knowledge of the Fermi
potential of the Fomblin grease. From the measured TOF spectrum shown in
fig. 6a we find $\left. \mathrm{d}\phi /\mathrm{d}\lambda \right\vert
_{\lambda ^{\ast }}=5.0\times 10^{8}$ cm$^{-2}$s$^{-1}$nm$^{-1}$, from which
we expect a single-phonon production rate density $P_{\mathrm{I}}=(0.61\pm
0.10)~$s$^{-1}$cm$^{-3}$.

The differential multi-phonon production rate density is given by

\begin{equation}
\frac{\mathrm{d}P_{\mathrm{II}}}{\mathrm{d}\lambda }=n_{^{4}\mathrm{He}%
}\sigma _{^{4}\mathrm{He}}E_{\mathrm{c}}\frac{k_{\mathrm{c}}}{3\pi }\frac{%
\mathrm{d}\phi }{\mathrm{d}\lambda }s_{\mathrm{II}}(\lambda )\lambda .
\end{equation}

\noindent $n_{^{4}\mathrm{He}}=$\ is the number density and $\sigma _{^{4}\mathrm{He}}=$ is the cross section per helium atom. $E_{\mathrm{c}}$ and $k_{\mathrm{c}}$
are the maximum kinetic energy and wavenumber of the neutrons trapped in the
vessel. $s_{\mathrm{II}}(\lambda )$ is the scattering function $s_{\mathrm{II}}(Q,\omega )$, evaluated on the dispersion curve of the free neutron, i.e. $\omega =\hbar Q^{2}/(2m_{\mathrm{n}})$, with $Q=2\pi /\lambda $. It is
modeled\footnote{More details about the modeling of $s_{\mathrm{II}}\left( \lambda \right) $
will be described in a forthcoming paper.} to fit data from \cite{Andersen/1994,Andersen/2007} which is also shown in fig. 6a. Integration of 
$dP_{\mathrm{II}}/d\lambda $ from $0.52$ nm to $\infty $ for the measured
incident differential flux provides an estimated lower limit of $0.42$ s$^{-1}$cm$^{-3}$ for the integral multi-phonon production rate $P_{\mathrm{II}}$. Difficult to estimate is the contribution to $P_{\mathrm{II}}$ from the
wavelength range $\lambda <0.52$ nm. First, there is no direct experimental
information available about $s_{\mathrm{II}}(\lambda )$ and second, this
region has a strong weight due to the relatively large $\mathrm{d}\phi /\mathrm{d}\lambda $. The $\lambda $-dependent measurements of UCN production
by Baker and colleagues \cite{Baker/2003} indicate that a multi-phonon
contribution exists also from the region $0.4$ $\mathrm{nm}\leq \lambda \leq 0.52$ nm. The measured integral multi-phonon and single-phonon production
rates were found to add up evenly to the measured integral production rate 
\cite{van der Grinten/2008}. This provides an experimental hint that for $\lambda <0.4$ nm the multi-phonon contribution might be negligible. The fit
function shown in fig. 6 was therefore designed to extend the range of
available data with a smooth decrease to zero at $0.4$ nm. Integration of $\mathrm{d}P_{\mathrm{II}}/\mathrm{d}\lambda $ using the complete fitting function for $s_{\mathrm{II}}(\lambda )$ yields $P_{\mathrm{II}}=0.56$ s$^{-1}$cm$^{-3}$.
Estimating the uncertainty of $P_{\mathrm{II}}$ as the difference between
the values with and without inclusion of the contribution from the range $0.4~\mathrm{nm}\leq \lambda \leq 0.52$ nm, we may expect a total UCN
production rate of

\begin{equation}
P_{\mathrm{th}}=(0.61\pm 0.10)_{\mathrm{I}}\text{ }\mathrm{s}^{-1}\mathrm{cm}
^{-3}+(0.56\pm 0.14)_{\mathrm{II}}\text{ }\mathrm{s}^{-1}\mathrm{cm}
^{-3}=(1.17\pm 0.17)\text{ }\mathrm{s}^{-1}\mathrm{cm}^{-3}.
\end{equation}

Evidently, more experimental information about $s_{\mathrm{II}}(\lambda )$
is required for a secure prediction of the UCN production rate induced by a
white cold neutron beam. Note that this is particularly necessary for the
region of short wavelengths, whereas the contribution to $P_{\mathrm{II}}$
from the region $\lambda >0.8$ nm is only $0.03$ s$^{-1}$cm$^{-3}$. We also
note that predictions from the two model calculations \cite%
{Korobkina/2002,Schott/2003} for the region $\lambda <0.52$ nm give mutually
inconsistent results.

\begin{figure}
	\centering
	\includegraphics[]{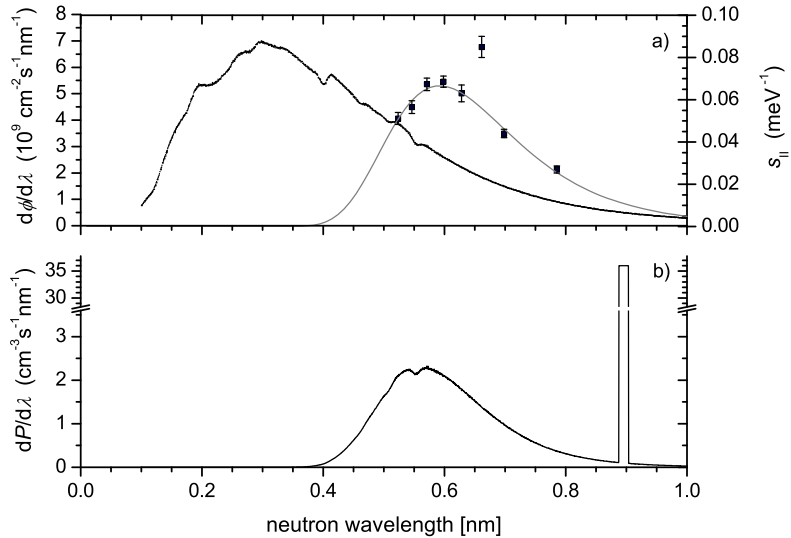}
\caption{a) Differential flux at the neutron guide NL1 of FRM\thinspace II
(left scale), and measured scattering function $s_{\mathrm{II}}(\protect \lambda )$ and fitted curve (right scale). The data point above the curve is
due to the roton-maxon resonance in the multi-phonon dynamic structure
factor. b) Calculated differential production rate, including single- and
multi-phonon contributions, employing the measured $\mathrm{d}\protect\phi /\mathrm{d}\protect\lambda $ and the fit function for $s_{\mathrm{II}}\left(\protect\lambda \right) $ from fig. 6 a). }
\end{figure}

\section{Discussion and conclusions}

Our experiments demonstrate the feasibility of a versatile, intense UCN
source at a cold neutron beam, providing UCN for experiments at room
temperature. With the present prototype, at $T=0.82$ K and using a small
extraction hole with area $A=2$ cm$^{2}$, about $60$ \% of the UCN produced
in the Fomblin grease-coated converter vessel reached the detector. Compared
to our previous experiment with the uncoated electropolished stainless steel
converter vessel \cite{Zimmer/2007}, the count rates of accumulated UCN
observed immediately after opening the UCN valve, and also the integral UCN
counts were almost a factor three higher, despite the lower Fermi potential
of the Fomblin wall coating. This increase of UCN output is due to longer
storage time constants and a faster passage of the UCN through the
extraction hole. Values for $\tau _{A}$ as derived from eq.(\ref{tau-e}) can
be compared to the gas-kinetic equation%

\begin{equation}
\tau _{A}=\frac{4V}{vA},  \label{tau-A}
\end{equation}

\noindent where we insert $V\approx 2.4$ l for the volume of the converter vessel, and
the mean velocity $v$ of the trapped UCN, which, for a uniform distribution
in velocity space, is given by $v=3v_{\max }/4$ with $m_{\mathrm{n}}v_{\max}^{2}/2=V_{\mathrm{F,wall}}-V_{\mathrm{F,}^{4}\mathrm{He}}$, where $m_{\mathrm{n}}$ is the neutron mass. From the data for the Fomblin-coated
vessel at $T=1.26$ K we thus expect $\tau _{A}=14.8$ s. The measured value, $ \tau _{A}=\left( 27\pm 3\right) $ s, is much closer to the expected value
than in our earlier experiments with the uncoated converter vessel. There,
from eq.(\ref{tau-A}) with the Fermi potential of stainless steel we may
expect $\tau _{A}=11.3$ s but measured $\tau _{A}=\left( 58\pm 13\right)$~s. A possible explanation is that UCN trajectories do not explore quickly
enough the available phase space in the electropolished vessel, a hypothesis
already raised earlier by W. Mampe and co-workers for their liquid-wall UCN
bottle experiment \cite{Mampe/1989}. Thus a certain roughness of the vessel
walls seems to be necessary for efficient extraction of the UCN through a
small hole. This should be taken into account in the design of a large
converter vessel in order to keep the time needed to extract the UCN
reasonably short.

The UCN production rate densities determined with two partly independent
methods agree very well, $\varepsilon P_{1}=\left( 0.80\pm 0.08\right) $~s$^{-1}$cm$^{-3}$, and $\varepsilon P_{2}=\left( 0.79\pm 0.09\right) $ s$^{-1}$%
cm$^{-3}$. In our earlier experiment with the uncoated stainless steel
vessel, we obtained the values $\varepsilon P_{1}=\left( 0.91\pm 0.21\right)$ s$^{-1}$cm$^{-3}$, and $\varepsilon P_{2}=\left( 1.09\pm 0.25\right)$~s$^{-1}$cm$^{-3}$. The accuracy of the comparison has thus been improved by
more than a factor two in the present experiments. More important, also the
comparison with the theoretical expectation has been improved by virtue of a
dedicated TOF analysis of the cold neutron beam spectrum, due to which we
may expect the UCN production rate density $P_{\mathrm{th}}=(1.17\pm 0.17)$ $\mathrm{s}^{-1}\mathrm{cm}^{-3}$, with single- and multi-phonon
contributions of similar size. Since the detector and extraction
efficiencies, expressed by the factor $\varepsilon $, cannot be expected as
perfect, the measured values come indeed very close to $P_{\mathrm{th}}$.

In the present experiments the saturated UCN density, normalised to the
production volume $V_{\mathrm{p}}$ did not exceed $46$ UCN per cm$^{3}$. To
avoid activation of the stainless steel vessel only a quarter of the total
volume was irradiated with the cold beam. Also for practical reasons,
significant divergence losses due to beam collimation were accepted. In
addition, for simplicity of these first tests, we did not yet employ a wall
material with low neutron absorption and a high Fermi potential of $250$ neV
or beyond, such as Be, BeO or diamond-like carbon \cite%
{Atchison/2006,Atchison/2006b}. There one will gain in UCN density both due
to the larger energy of the storable neutrons and due to a larger storage
time. In order to produce a larger total number of UCN, the converter vessel
can be made much longer, due to the low cross section of $27\times 10^{-27}$
cm$^{2}$ for neutrons with wavelength $0.89$ nm \cite{Sommers/1955},
corresponding to a mean free path of about $17$ m. With a properly designed $ ^{4}$He converter and using an existing intense cold neutron beam at a high
flux reactor, UCN densities in the order $10^{4}$ per cm$^{3}$ with a total
UCN number up to several $10^{8}$ seem within reach.\medskip

\textbf{Acknowledgements: }We are very grateful to the head of the Physics
Department E18, Prof. S. Paul, and to the scientific director of the FRM II,
Prof. W. Petry, for supporting this development. We also thank I. Altarev,
A. M\"{u}ller and W. Schott for some useful discussions. We gratefully
acknowledge the help of M. Pfaller and his team from the central mechanical
workshop of the physics faculty for careful manufacturing of many cryostat
components. This work has been funded by the German BMBF (contract number
06MT250).

\end{document}